\documentclass[aps,prb,showpacs,groupedaddress,floatfix,nofootinbib]{revtex4}
\usepackage{amsmath}
\usepackage{graphicx}
\usepackage{color}

\begin{document}

\title{Simple one--dimensional quantum--mechanical model for a particle attached to
a surface}
\author{Francisco M. Fern\'andez}\email{fernande@quimica.unlp.edu.ar}

\affiliation{INIFTA (UNLP, CCT La Plata--CONICET), Divisi\'on Qu\'imica
Te\'orica \\Blvd. 113 y 64 S/N,
Sucursal 4, Casilla de Correo 16, 1900 La Plata, Argentina}

\begin{abstract}
We present a simple one--dimensional quantum--mechanical model for a
particle attached to a surface. We solve the Schr\"{o}dinger equation in
terms of Weber functions and discuss the behavior of the eigenvalues and
eigenfunctions. We derive the virial theorem and other exact relationships
as well as the asymptotic behaviour of the eigenvalues. We calculate the
zero--point energy for model parameters corresponding to H adsorbed on
Pd(100) and also outline the application of the Rayleigh--Ritz variational
method.
\end{abstract}


\maketitle

\section{Introduction}

\label{sec:intro}

In an introductory course on quantum theory one commonly discusses some of
the simplest models, such as, for example: free particle, particle in a box,
harmonic oscillator, tunnelling through a square barrier, etc. with the
purpose of making the students more familiar with the principles or
postulates of quantum theory. Some time ago Gibbs\cite{G75} introduced the
quantum bouncer as a model for the pedagogical discussion of some of the
relevant features of quantum theory. He derived the solutions to the
Schr\"{o}dinger equation in terms of Airy functions and obtained the energy
spectrum for two different model settings.

A closely related model, the harmonic oscillator with a hard wall on one
side, had been discussed earlier by Dean\cite{D66} and later by Mei and Lee%
\cite{ML83}. This model is as simple as the quantum bouncer and both can
therefore be discussed in the same course. An interesting feature of this
model is that it may in principle be useful to simulate a particle attached
to a wall like, for example, an atom adsorbed on a solid surface\cite
{RPG69,GKKSSO10}. Although it is an oversimplified one--dimensional model of
the actual physical phenomenon we deem it worthwhile to discuss some of its
properties in this paper.

In Sec.~\ref{sec:model} we introduce the model, write the Schr\"{o}dinger
equation in dimensionless form, and discuss some of the properties of its
solutions. In Sec.~\ref{sec:results} we obtain the eigenfunctions and
eigenvalues explicitly in terms of the well known Weber functions\cite
{MF53} and show the behavior of the eigenvalues and excitation energies
with respect to the distance between the particle and the wall. We also
calculate the zero--point energy for values of the model parameters
corresponding to de adsorption of H on Pd(100)\cite{GKKSSO10} and outline
the application of the Rayleigh--Ritz variational method to the Schr\"{o}%
dinger equation. Finally, is Sec.~\ref{sec:comments} we give further reasons
why this model may be useful in a course on quantum theory.

\section{Simple model}

\label{sec:model}

We consider a simple one--dimensional model for a particle of mass $m$
attached to a surface located at $x=0$ that separates free space ($x>0$)
from the bulk of the material ($x<0$). Therefore, we assume that the
potential exhibits an attractive tail for $x>0$ that keeps the particle in
the neighborhood of the surface and a repulsive one for $x<0$ that prevents
the particle from penetrating too deep into the material. Since we want to
keep the model as simple as possible we choose
\begin{equation}
V(x)=\left\{
\begin{array}{c}
\infty \text{ if }x<0 \\
\frac{k}{2}(x-d)^{2}\text{ if }x\geq 0
\end{array}
\right.  \label{eq:V(x)}
\end{equation}
where $k,d>0$. We note that the particle oscillates about $x=d$ but the
motion is not harmonic because of the effect of the hard wall.

The Schr\"{o}dinger equation reads
\begin{eqnarray}
&-&\frac{\hbar ^{2}}{2m}\psi ^{\prime \prime }(x)+\frac{k}{2}(x-d)^{2}\psi
(x)=E\psi (x)  \nonumber \\
&&\psi (0)=0,\,\lim_{x\rightarrow \infty }\psi (x)=0  \label{eq:Schr_model}
\end{eqnarray}
where the boundary condition at $x=0$ comes from the fact that in this
simple model the particle cannot penetrate into the material and,
consequently, $\psi (x)=0$ for all $x<0$. Since it is more convenient to
work with a dimensionless equation we define the length unit $L=\left( \frac{%
\hbar ^{2}}{km}\right) ^{1/4}$ and the dimensionless coordinate $q=(x-d)/L$.
Thus, the dimensionless Schr\"{o}dinger equation reads
\begin{eqnarray}
&-&\frac{1}{2}\varphi ^{\prime \prime }(q)+\frac{1}{2}q^{2}\varphi
(q)=\epsilon \varphi (q)  \nonumber \\
&&\varphi (-q_{0})=0,\,\lim_{q\rightarrow \infty }\varphi (q)=0
\label{eq:Schr_dim}
\end{eqnarray}
where
\begin{eqnarray}
&&\epsilon =\frac{mL^{2}E}{\hbar ^{2}}=\frac{E}{\hbar \omega },\text{%
\thinspace }\omega =\sqrt{\frac{k}{m}}  \nonumber \\
&&q_{0}=\frac{d}{L}=d\left( \frac{km}{\hbar ^{2}}\right) ^{1/4}  \nonumber \\
&&\varphi (q)=\psi (Lq+d)  \label{eq:mod_param}
\end{eqnarray}
It is worth noting that $q_{0}$ increases with $d$, $m$, and $k$.
It is precisely Eq. (\ref{eq:Schr_dim}) that was discussed by Dean\cite{D66}
and Mei and Lee\cite{ML83}.

When $q_{0}\rightarrow \infty $ we have the well--known harmonic oscillator
with eigenvalues
\begin{equation}
\lim_{q_{0}\rightarrow \infty }\epsilon _{n}=n+\frac{1}{2},\,n=0,1,\ldots
\label{eq:en_free}
\end{equation}
On the other hand, when $q_{0}=0$ we have the harmonic oscillator in the
half line and
\begin{equation}
\lim_{q_{0}\rightarrow 0}\epsilon _{n}=2n+\frac{3}{2},\,n=0,1,\ldots
\label{eq:en_half}
\end{equation}
Note that these are merely the harmonic--oscillator eigenvalues with odd
quantum number (the corresponding eigenfunctions have a node at origin). It
follows from Eq.~(\ref{eq:de/db_2}) in the Appendix that
\begin{equation}
\frac{d\epsilon }{dq_{0}}<0,\;\lim_{q_{0}\rightarrow \infty }\frac{d\epsilon
}{dq_{0}}=0  \label{eq:de/dq0}
\end{equation}
from which we conclude that the energy eigenvalues decrease monotonously
between the following limits:
\begin{equation}
2n+\frac{3}{2}\geq \epsilon _{n}(q_{0}) > n+\frac{1}{2},\,n=0,1,\ldots
\label{eq:en_bounds}
\end{equation}
when $0 \leq q_{0}<\infty $. It should be kept in mind that the number of zeros
of $\varphi _{n}(q)$ remains unchanged as $q_{0}$ goes from $0$ to $\infty $.

Equation (\ref{eq:de/db_2}) is useful for obtaining the asymptotic behavior
of the energy as $q_{0}\rightarrow \infty $. To this end we simply integrate
it from $-\infty $ to $-q_{0}$:
\begin{equation}
\epsilon (-q_{0})=\epsilon (-\infty )+\frac{1}{2}\int_{-\infty }^{-q_{0}}%
\frac{\varphi ^{\prime }(b)^{2}}{\int_{b}^{\infty }\varphi (q)^{2}\,dq}\,db
\label{eq:e(-q0)}
\end{equation}
For the ground state we expect that $\varphi (q)\rightarrow Ne^{-q^{2}/2}$
as $q_{0}\rightarrow \infty $. The normalization factor is approximately
given by
\begin{equation}
N^{-2}(q_{0})=\int_{-q_{0}}^{\infty }e^{-q^{2}}\,dq=\frac{\sqrt{\pi }}{2}[1+%
\mbox{erf}(q_{0})]
\end{equation}
where $\mbox{erf}(z)$ is the error function. Since $\mbox{erf}(z)\leq %
\mbox{erf}(\infty )=1$ we write $\mbox{erf}(q_{0})=1-\xi $ and expand $N^{2}$
in a Taylor series about $\xi =0$:
\begin{equation}
N(q_{0})^{2}\approx \frac{2+\xi }{2\sqrt{\pi }}=\frac{3-\mbox{erf}(q_{0})}{2%
\sqrt{\pi }}
\end{equation}
Finally, Eq.~(\ref{eq:e(-q0)}) yields
\begin{equation}
\epsilon _{0}(-q_{0})\approx \frac{1}{2}+\frac{q_{0}e^{-q_{0}^{2}}}{2\sqrt{%
\pi }}  \label{eq:e0(-q0)}
\end{equation}
In order to obtain this result we substituted $\mbox{erf}(q_{0})\approx 1$
after the integration and neglected a term proportional to $e^{-2q_{0}^{2}}$
because it is much smaller than the exponential one retained in Eq.~(\ref
{eq:e0(-q0)}). This result agrees with the one derived by Mei and Lee\cite
{ML83} by means of perturbation theory (note that their parameter $R$ is $%
\sqrt{2}q_{0}$).

Proceeding in the same way for the first excited state we obtain
\begin{equation}
\epsilon _{1}(-q_{0})\approx \frac{3}{2}+\frac{%
q_{0}(2q_{0}^{2}-1)e^{-q_{0}^{2}}}{2\sqrt{\pi }}  \label{eq:e1(-q0)}
\end{equation}
that is slightly different from the result of Mei and Lee\cite{ML83}.
However, they are equivalent for most purposes because the difference
between them is smaller than their absolute errors.

Eq.~(\ref{eq:virial_2}) gives us the virial theorem\cite{FC87} for this model
\begin{equation}
\left\langle \hat{D}^{2}\right\rangle +\left\langle q^{2}\right\rangle =q_{0}%
\frac{\partial \epsilon }{\partial q_{0}}  \label{eq:virial_HO}
\end{equation}
where $\hat{D}=d/dq$. The right--hand side of this equation is the virial of
the force exerted by the surface. Eq.~(\ref{eq:HV2}) provides us with
another interesting relation
\begin{equation}
\left\langle q\right\rangle =-\frac{\partial \epsilon }{\partial q_{0}}
\label{eq:hypervirial_HO}
\end{equation}
that clearly reveals the asymmetry of the interaction between the particle
and the surface. In both cases we recover the well--known results for the
harmonic oscillator $\left\langle \hat{D}^{2}\right\rangle +\left\langle
q^{2}\right\rangle =0$ and $\left\langle q\right\rangle =0$ when $%
q_{0}\rightarrow \infty $.

\section{Results}

\label{sec:results}

If we define the new independent variable $z=\sqrt{2}q$ and write the energy
as $\epsilon =m+1/2$ then we realize that $D_{m}(z)=\varphi (z/\sqrt{2})$ is
a solution to the Weber equation\cite{MF53}
\begin{equation}
D_{m}^{\prime \prime }(z)+\left( m+1/2-z^{2}/4\right) D_{m}(z)=0
\label{eq:Weber_eq}
\end{equation}
The general solution is\cite{MF53}
\begin{equation}
D_{m}(z)=2^{m/2}\sqrt{\pi }e^{-z^{2}/4}\left[ \frac{1}{\Gamma (\frac{1-m}{2})%
}F\left( -\frac{m}{2}\left| \frac{1}{2}\right| \frac{z^{2}}{2}\right) -\frac{%
\sqrt{2}z}{\Gamma (-\frac{m}{2})}F\left( \frac{1-m}{2}\left| \frac{3}{2}%
\right| \frac{z^{2}}{2}\right) \right]  \label{eq:Weber_func}
\end{equation}
where the confluent hypergeometric function $F\left( a\left| c\right|
z\right) $ is a solution to
\begin{equation}
zF^{\prime \prime }(z)+(c-z)F(z)-aF(z)=0  \label{eq:hypergeom_eq}
\end{equation}
and can be expanded in a Taylor series about $z=0$ as
\begin{equation}
F\left( a\left| c\right| z\right) =1+\frac{a}{c}z+\frac{a(a+1)}{2!c(c+1)}%
z^{2}+\ldots  \label{eq:hypergeom_func}
\end{equation}

The boundary condition $\varphi (-q_{0})=0$ enables us to calculate the
eigenvalues from the roots of $D_{m}\left( -\sqrt{2}q_{0}\right) =0$. For
each value of $q_{0}$ we solve
\begin{equation}
\frac{1}{\Gamma (\frac{1-m}{2})}F\left( -\frac{m}{2}\left| \frac{1}{2}%
\right| q_{0}^{2}\right) +\frac{2q_{0}}{\Gamma (-\frac{m}{2})}F\left( \frac{%
1-m}{2}\left| \frac{3}{2}\right| q_{0}^{2}\right) =0  \label{eq:En_equation}
\end{equation}
for $m$ and then calculate the dimensionless energy $\epsilon =m+1/2$. This
approach has already been discussed by Dean\cite{D66} and Mei and Lee\cite
{ML83}.

\begin{figure}[H]
\begin{center}
\includegraphics[width=9cm]{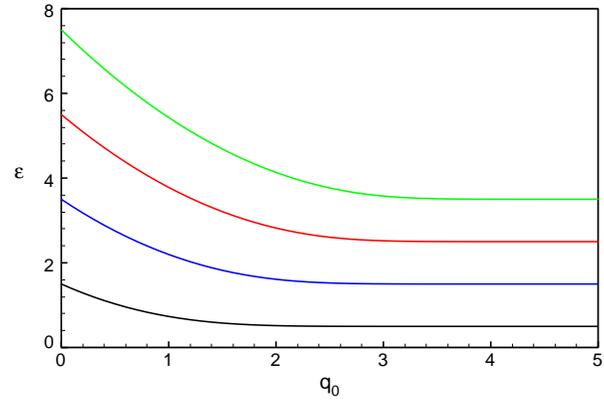}
\end{center}
\caption{(Color online) First four energy levels}
\label{fig:en}
\end{figure}

Fig.~\ref{fig:en} shows $\epsilon _{n}(q_{0})$ for $n=0,1,2,3$ and a wide
range of values of $q_{0}$. We note that the dimensionless energy decreases
monotonously as predicted by Eq. (\ref{eq:de/dq0}) between the limits
indicated in Eq. (\ref{eq:en_bounds}).

\begin{figure}[H]
\begin{center}
\includegraphics[width=9cm]{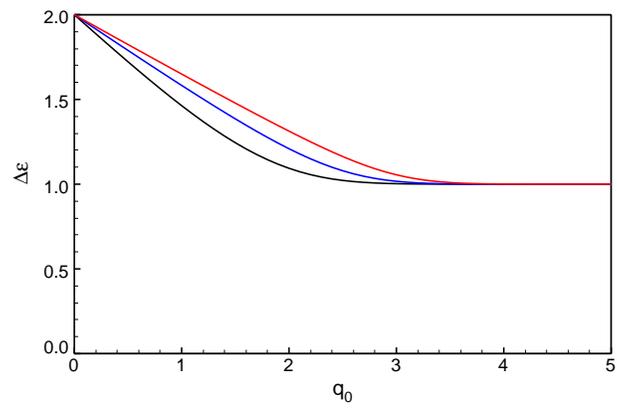}
\end{center}
\caption{(Color online) The first three excitation energies
$\epsilon_{n+1}-\epsilon_n$ ($n=0,1,2$ from bottom to top)}
\label{fig:De}
\end{figure}

The gap between two consecutive energy levels of the harmonic oscillator is $%
\Delta \epsilon _{n}=\epsilon _{n+1}-\epsilon _{n}=1$ ($\Delta E_{n}=\hbar
\omega $). Fig.~\ref{fig:De} shows $\Delta \epsilon _{n}(q_{0})$ for
$n=0,1,2 $ where we note that the energy gap increases with $n$ revealing
that the presence of the wall results in an anharmonic oscillation. It
becomes more harmonic as $q_{0}$ increases (by increasing either
$d$, $m$ or $k$).

In order to have a clearer physical idea of the kind of predictions of this
simple model we may choose the parameters for the adsorption of hydrogen on
Pd(100). Gladys et al\cite{GKKSSO10} estimated $d\approx 0.4\,$\AA ~and
$k\approx 15\,N\,m^{-1}$ for hydrogen that lead to $q_{0}\approx 1.55$.
The zero--point energy for such model parameters is
approximately $0.57\,\hbar \,\omega $ instead of the value $\hbar \,\omega /2
$ chosen by those authors. Although they fitted the potential energy of the
vertical displacement of the H atom from the Pd surface to a cubic
polynomial they simply chose the zero--point energy of the harmonic
oscillator. Present model predicts that the zero--point energy is slightly
greater than the harmonic--oscillator one because of the repulsive effect of
the surface. We agree that a hard wall may not be the most adequate
representation of the short--range interaction between the H atom and the Pd
surface, but we think that the model is a reasonably simple first approach
to the physical phenomenon.

For deuterium we have $d\approx 0.45\,$\AA ~which, together with the same
force constant and about twice the mass, yields $q_{0}\approx 2$
and $\epsilon
_{0}\approx 0.52\,\hbar \,\omega $ that is closer to the
harmonic--oscillator zero--point energy\cite{GKKSSO10}. We note the effect of
the distance
to the surface and the mass of the particle on the vibrational energies.

This model is also useful for discussing the Rayleigh--Ritz variational
method\cite{MF53}. If, for example, we choose the trial function
\begin{equation}
u(q)=\sum_{j=0}^{N-1}c_{j}f_{j}(q)  \label{eq:u_trial}
\end{equation}
as a linear combination of the non orthogonal basis set
\begin{equation}
f_{j}(q)=(q+q_{0})q^{j}e^{-q^{2}/2},\;j=0,1,\ldots   \label{eq:f_j}
\end{equation}
and minimize the approximate energy
\begin{eqnarray}
w &=&\frac{\int_{-q_{0}}^{\infty }u(q)\hat{H}u(q)\,dq}{\int_{-q_{0}}^{\infty
}u(q)^{2}\,dq}  \nonumber \\
\hat{H} &=&-\frac{1}{2}\frac{d^{2}}{dq^{2}}+\frac{1}{2}q^{2}
\label{eq:w_var}
\end{eqnarray}
then we arrive at the secular equations
\begin{eqnarray}
&&\sum_{j=0}^{N-1}\left( H_{ij}-wS_{ij}\right) c_{j}=0,\;i=0,1,\ldots ,N-1
\nonumber \\
&&H_{ij}=\int_{-q_{0}}^{\infty }f_{i}(q)\hat{H}f_{j}(q)\,dq,\;S_{ij}=%
\int_{-q_{0}}^{\infty }f_{i}(q)f_{j}(q)\,dq  \label{eq:secular_eqs}
\end{eqnarray}
There are nontrivial solutions for the coefficients $c_{j}$ only if the
secular determinant vanishes
\begin{equation}
\left| H_{ij}-wS_{ij}\right| _{i,j=0}^{N-1}=0  \label{eq:secular_det}
\end{equation}
Its roots $w_{n}^{[N]}$, $n=0,1,\ldots ,N-1$ are upper bounds to the actual
eigenvalues and satisfy
\begin{equation}
w_{n}^{[N]}>w_{n}^{[N+1]}>\epsilon _{n}  \label{eq:var_bounds}
\end{equation}
The particle attached to a wall is therefore a suitable example for
illustrating how the roots of the secular determinant approach the
eigenvalues (calculated accurately from the Weber functions) from above as $N
$ increases. We do not show results here and simply mention that the
calculation is greatly facilitated by the fact that one calculates the
integrals $H_{ij}$ and $S_{ij}$ analytically.

\section{Further comments}

\label{sec:comments}

The model discussed here is suitable for a course on quantum theory because
it does not require much more mathematical background than it is necessary
for the discussion of the well known harmonic oscillator or the quantum
bouncer\cite{G75}. It is useful for introducing a numerical calculation of
the eigenvalues that the student does not find in the treatment of the
harmonic oscillator. One can approach the problem by means of either the Weber
functions or the Rayleigh--Ritz variational method. The student
will also learn that it is necessary to modify the form of the well known virial
theorem and other mathematical expressions in order to take into account the
effect of the wall. We believe that present derivation of the analytical
results in Sec.~\ref{sec:model} and in the Appendix is
simpler than those available in the scientific literature\cite{D66,ML83}.

In addition to it, the model enables us to simulate the adsorption of an
atom on a surface and discuss anharmonic vibrations in quantum theory. In
the study of molecular vibrations one introduces nonlinear
oscillations by means of cubic, quartic, and other terms of greater degree
in the potential--energy function. In this case it arises from the boundary
condition forced by the hard wall.

\appendix*
\section{Some useful mathematical relations}

\label{sec:appendix}

In this appendix we develop some useful analytical results for the
eigenfunctions and eigenvalues of the constrained oscillator. Although
similar expressions have already been shown elsewhere\cite{FC87} we derive
them here in a form that is more suitable for our needs.

Consider the dimensionless Schr\"{o}dinger equation
\begin{eqnarray}
&-&\frac{1}{2}\varphi ^{\prime \prime }(x)+V(x)\varphi (x)=\epsilon \varphi
(x)  \nonumber \\
&&\varphi (b)=0,\,\lim_{x\rightarrow \infty }\varphi (x)=0
\label{eq:Schr_gen}
\end{eqnarray}
Both the eigenvalue $\epsilon $ and the eigenfunction $\varphi (x)$ depend
on the chosen value of $b$. If we differentiate this equation with respect
to $b$ and call $\chi (x)=\partial \varphi (x)/\partial b$ we have
\begin{equation}
-\frac{1}{2}\chi ^{\prime \prime }(x)+V(x)\chi (x)=\epsilon \chi (x)+\frac{%
d\epsilon }{db}\varphi (x)  \label{eq:dSchro/db}
\end{equation}
If we multiply this equation by $\varphi (x)$ and integrate the result
between $b$ and $\infty $ we easily obtain
\begin{equation}
\frac{d\epsilon }{db}\int_{b}^{\infty }\varphi (x)^{2}\,dx=-\frac{1}{2}\chi
(b)\varphi ^{\prime }(b)  \label{eq:de/db_1}
\end{equation}
because the integration by parts of the first term yields
\begin{equation}
\int_{b}^{\infty }\chi ^{\prime \prime }(x)\varphi (x)\,dx=\chi (b)\varphi
^{\prime }(b)+2\int_{b}^{\infty }\left[ V(x)-\epsilon \right] \chi
(x)\varphi (x)\,dx
\end{equation}

If we now differentiate the boundary condition $\varphi (b)=0$ and take into
account that $\varphi (x)$ depends on $b$ we obtain
\begin{equation}
\chi (b)+\varphi ^{\prime }(b)=0  \label{eq:dphi/db}
\end{equation}
so that equation (\ref{eq:de/db_1}) becomes
\begin{equation}
\frac{d\epsilon }{db}\int_{b}^{\infty }\varphi (x)^{2}\,dx=\frac{1}{2}%
\varphi ^{\prime }(b)^{2}  \label{eq:de/db_2}
\end{equation}

We define the operators
\begin{equation}
\hat{H}=-\frac{1}{2}\hat{D}^{2}+V(x)  \label{eq:H_op}
\end{equation}
and $\hat{v}=\hat{x}\hat{D}$, where $\hat{D}=d/dx$. The commutator between
them reads
\begin{equation}
\lbrack \hat{H},\hat{v}]=\hat{H}\hat{v}-\hat{v}\hat{H}=-\hat{D}%
^{2}-xV^{\prime }
\end{equation}
If $\varphi (x)$ is an eigenfunction of $\hat{H}$ with eigenvalue $\epsilon $
then straightforward integration by parts leads to
\begin{equation}
\int_{b}^{\infty }\varphi [\hat{H},\hat{v}]\varphi \,dx=-\frac{1}{2}b\varphi
^{\prime }(b)^{2}  \label{eq:virial_1}
\end{equation}
which, by virtue of Eq.~(\ref{eq:de/db_2}), becomes the virial theorem
\begin{equation}
\left\langle \hat{D}^{2}\right\rangle +\left\langle xV^{\prime
}\right\rangle =b\frac{\partial \epsilon }{\partial b}  \label{eq:virial_2}
\end{equation}
where
\begin{equation}
\left\langle \hat{A}\right\rangle =\frac{\int_{b}^{\infty }\varphi \hat{A}%
\varphi \,dx}{\int_{b}^{\infty }\varphi ^{2}\,dx}
\end{equation}

Analogously, from the commutator $[\hat{H},\hat{D}]=-V^{\prime }$ we obtain
\begin{equation}
\int_{b}^{\infty }\varphi (x)^{2}V^{\prime }(x)\,dx=\frac{1}{2}\varphi
^{\prime }(b)^{2}  \label{eq:HV1}
\end{equation}
or
\begin{equation}
\left\langle V^{\prime }\right\rangle =\frac{\partial \epsilon }{\partial b}
\label{eq:HV2}
\end{equation}

We can test these equations quite easily by means of the well--known
solutions to the free harmonic oscillator
\begin{equation}
\psi _{n}=N_{n}H_{n}(x)e^{-x^{2}/2}  \label{eq:HO-Psi_n}
\end{equation}
where $H_{n}(x)$ is a Hermite polynomial and $N_{n}$ the corresponding
normalization factor\cite{MF53}. If we choose $b$ to be one of the zeroes $%
x_{nj}$, $j=1,2,\ldots ,n$ of $H_{n}(x)$, $n=1,2,\ldots $ then we verify that $%
\psi _{n}(x)$ satisfies equations (\ref{eq:virial_1}) and (\ref{eq:HV1}) for
$V(x)=x^{2}/2$. If, for example, $b=0$ for $n=1$ and $b=\pm 1/\sqrt{2}$ for
$n=2 $ then $\epsilon =3/2$ is the energy of the ground state of the harmonic
oscillator with the boundary condition at $b=0$ and $\epsilon =5/2$ is the
energy of the ground state with $b=1/\sqrt{2}$ and of the first excited
state with $b=-1/\sqrt{2}$.

\end{document}